\documentclass{article}

\usepackage{natbib}
\usepackage[letterpaper,top=2cm,bottom=2cm,left=3cm,right=3cm,marginparwidth=1.75cm]{geometry}

\usepackage{amsmath}
\usepackage{amssymb}
\usepackage{amsthm}
\usepackage{graphicx}
\usepackage[colorlinks=true, allcolors=blue]{hyperref}

\title{\textbf{Predicting Stock Market Crash with Bayesian Generalised Pareto Regression}}
\author{
    Sourish Das\footnote{corresponding author: sourish@cmi.ac.in}
}

\date{
    June 19, 2025
}

\begin{document}
\maketitle

%%%%%%%%%%%%%%%%%%%%% ABSTRACT %%%%%%%%%%%%%%%%%%%%%%%%%%%%%%%%
\begin{abstract}
\noindent This paper develops a Bayesian Generalised Pareto Regression (GPR) model to forecast extreme losses in Indian equity markets, with a focus on the Nifty 50 index. Extreme negative returns, though rare, can cause significant financial disruption, and accurate modelling of such events is essential for effective risk management. Traditional Generalised Pareto Distribution (GPD) models often ignore market conditions; in contrast, our framework links the scale parameter to covariates using a log-linear function, allowing tail risk to respond dynamically to market volatility. We examine four prior choices for Bayesian regularisation of regression coefficients: Cauchy, Lasso (Laplace), Ridge (Gaussian), and Zellner’s $g$-prior. Simulation results suggest that the Cauchy prior delivers the best trade-off between predictive accuracy and model simplicity, achieving the lowest RMSE, AIC, and BIC values. Empirically, we apply the model to large negative returns (exceeding 5\%) in the Nifty 50 index. Volatility measures from the Nifty 50, S\&P 500, and gold are used as covariates to capture both domestic and global risk drivers. Our findings show that tail risk increases significantly with higher market volatility. In particular, both S\&P 500 and gold volatilities contribute meaningfully to crash prediction, highlighting global spillover and flight-to-safety effects. The proposed GPR model offers a robust and interpretable approach for tail risk forecasting in emerging markets. It improves upon traditional EVT-based models by incorporating real-time financial indicators, making it useful for practitioners, policymakers, and financial regulators concerned with systemic risk and stress testing.
\end{abstract}

%%%%%%%%%%%%%%%%%%%%% KEYWORDS %%%%%%%%%%%%%%%%%%%%%%%%%%%%%%%%

%%%%%%%%%%%%%%%%% AMS SUBJECT CLASSIFICATIONS %%%%%%%%%%%%%%%%%
 \vspace{-5mm} % This is optional

%%%%%%%%%%%%%%%%%%%%%%%% SECTION 1  %%%%%%%%%%%%%%%%%%%%%%%%%%%
\section{Introduction}

Stock market crashes have significant economic implications, often resulting in large wealth erosion, investor panic, and long-lasting financial instability, see \cite{liu2021impact, song20222020}. Understanding and anticipating such extreme events is vital for effective risk management, financial regulation, and economic forecasting, see \cite{dai2021preventing}. During the COVID-19 crash, for instance, \citet{giglio2020inside} document a sharp decline in short-term investor expectations and a surge in disagreement about future market performance. \citet{mazur2021covid} further highlight that while sectors like healthcare and software showed resilience, others such as hospitality and energy experienced extreme negative returns and asymmetric volatility, accompanied by varied corporate responses.

Extreme Value Theory (EVT) provides a principled framework for modelling the statistical behaviour of rare but severe events, particularly through the Generalised Pareto Distribution (GPD), which is commonly used for modelling exceedances over a high threshold, see \cite{smith1985mle}. In the context of financial crashes, \citet{fry2008statistical} presents a comprehensive investigation into bubbles, volatility, and contagion, deriving a GPD-based model for market drawdowns to analyse tail risk dynamics.

Several studies have applied the GPD to model extreme financial phenomena. \citet{malevergne2006power} and \citet{das2016understanding} use EVT to characterise financial extremes, but their models lack covariate inputs, limiting their capacity to account for underlying market conditions such as volatility or liquidity. \citet{liu2011detecting} advances the field by detecting structural breaks in tail behaviour using a transformed GPD and fluctuation tests, highlighting the need for flexible models when estimating extreme quantiles under changing market regimes.

More recent work by \citet{rai2022statistical} analyses the statistical behaviour of aftershocks in stock market crashes and finds that tail patterns and inter-occurrence times vary depending on the nature of the crash. This supports the hypothesis that covariate information, such as macroeconomic signals, volatility, or liquidity, can influence the shape and scale of extreme return distributions.

Motivated by these findings, we extend the standard GPD framework by modelling the scale parameter as a function of covariates through a log-linear link. This Generalised Pareto Regression (GPR) model enhances both interpretability and predictive capacity by incorporating market-relevant features directly into the tail distribution.

Earlier examples of GPR models include \citet{das2010analysis}, who model extreme alcohol consumption events using a covariate-linked scale parameter within a GPD framework. In the operational risk domain, \citet{hambuckers2018understanding} employ a regularised GPD regression where both the scale and shape parameters vary with covariates. While flexible, such models risk overparameterisation and identifiability issues. To maintain parsimony and model explainability, we assume a common shape parameter across observations and model only the scale parameter as a function of predictors.

Next in Section \ref{Sec_Data} we present the data description and exploratory data analysis. In Section \ref{Sec_Method} we present the Bayesian methodology for Generalised Pareto Regression. In Section \ref{Sec_Simulation}, we present a thorough simulation study to evaluate the performance of different methodologies. In Section \ref{Sec_Empirical}, we present the Bayesian analysis of Indian market crash events. Section \ref{Sec:Conclusion} conclude the paper.

\section{Data Insight}\label{Sec_Data}

\begin{figure}[ht]
    \centering
    \includegraphics[width=0.22\linewidth]{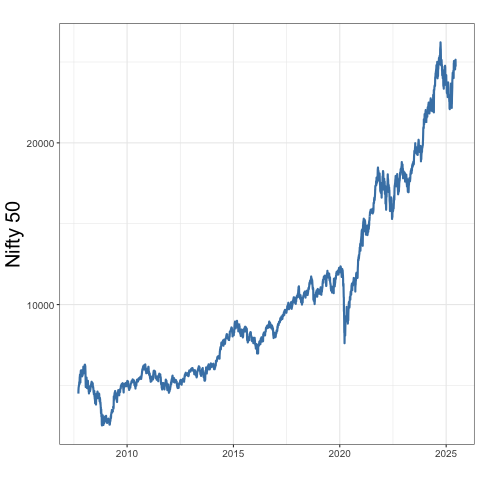}
    \includegraphics[width=0.22\linewidth]{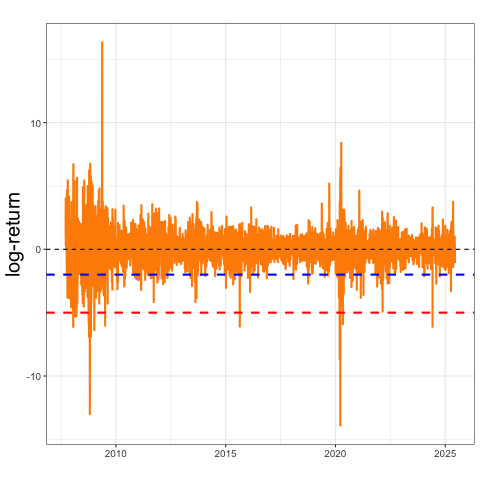}
    \includegraphics[width=0.22\linewidth]{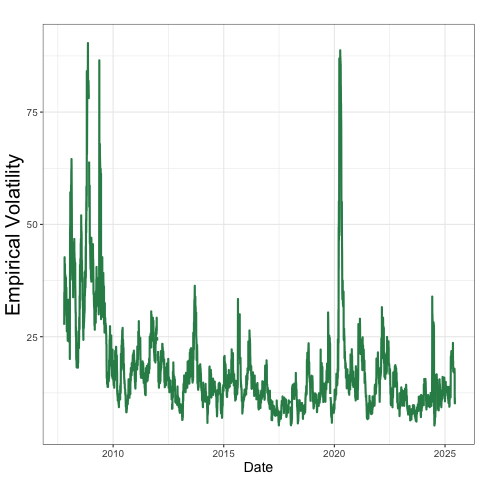}
    \includegraphics[width=0.22\linewidth]{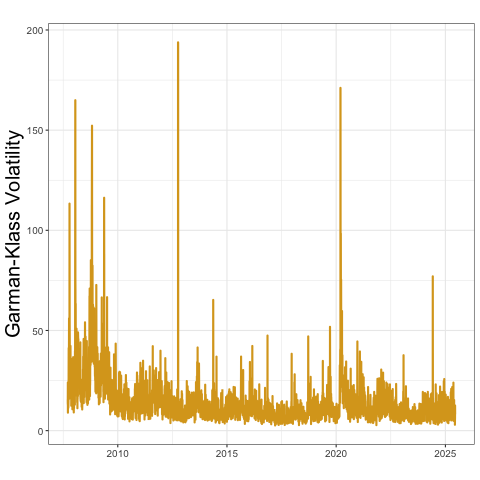}\\
    \small
    \parbox{0.22\linewidth}{\centering (a)}%
    \parbox{0.22\linewidth}{\centering (b)}%
    \parbox{0.22\linewidth}{\centering (c)}%
    \parbox{0.22\linewidth}{\centering (d)}\\[1ex]

    \includegraphics[width=0.32\linewidth]{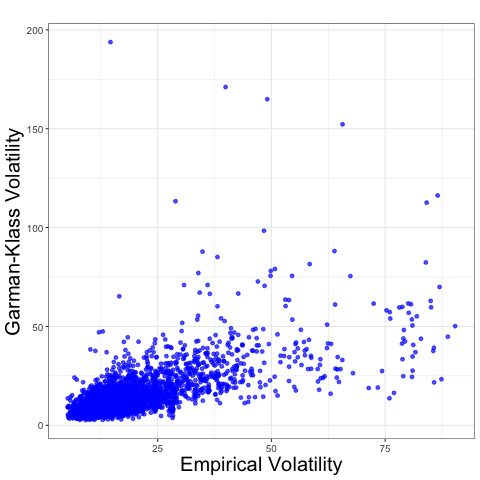}
    \includegraphics[width=0.32\linewidth]{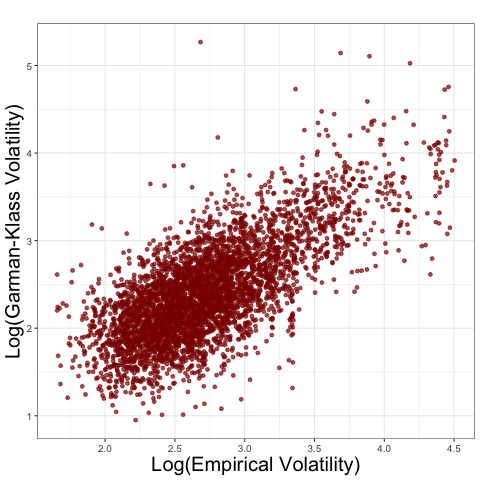}\\
    \small
    \parbox{0.32\linewidth}{\centering (e)}%
    \parbox{0.32\linewidth}{\centering (f)}

    \caption{Visual summary of Nifty50 prices and volatility measures: (a) closing price, (b) log-return, (c) empirical volatility, (d) Garman-Klass volatility, (e) empirical vs GK volatility, and (f) log-transformed comparison of empirical vs GK volatility.}
    \label{fig:nifty-panel}
\end{figure}

We analyse the Nifty 50 index over the period from 17 September 2007 to 13 June 2025. Figure~\ref{fig:nifty-panel}(a) shows the daily closing prices of the Nifty 50, and Figure~\ref{fig:nifty-panel}(b) displays the corresponding log-returns. The blue dashed line marks the $-2\%$ threshold and the red dashed line indicates the $-5\%$ threshold in returns. During this period, the index experienced a few sharp declines exceeding $5\%$, most notably during the 2008 global financial crisis and the COVID-19 pandemic. Given that the daily volatility of log-returns is approximately $1.32\%$, a drop of more than $2\%$ is typically regarded as a significant shock. A decline exceeding $5\%$ in a single day can trigger widespread panic, often resulting in substantial margin calls in the derivatives market, particularly in futures and options, which may lead to forced liquidations and amplified market instability.

Figure~\ref{fig:nifty-panel}(c) displays the empirical volatility of daily returns. This is computed using an exponentially weighted moving average (EWMA) method that considers past return variance over a rolling window of $k = 21$ trading days. Specifically, the conditional variance at time $t$ is calculated as
\[
    \sigma_t^2 = \alpha \cdot s_{t-1}^2 + (1 - \alpha) \cdot r_t^2,
\]
where $r_t$ is the daily return, $s_{t-1}^2$ is the sample variance of returns over the previous $k - 1$ days, and $\alpha = 0.9$ is the smoothing parameter; see \cite{sen2023computational}. The resulting daily volatility is annualised by multiplying by $\sqrt{250}$, reflecting the approximate number of trading days in a year. This approach yields a more adaptive and responsive measure of recent market volatility compared to traditional historical estimates.

Figure~\ref{fig:nifty-panel}(d) presents the Garman-Klass estimate of intraday volatility. This estimator offers a more efficient measure of daily volatility by incorporating high, low, opening, and closing prices, see \cite{garman1980estimation}. It reduces estimation variance by using price range data instead of just closing prices. The Garman-Klass volatility is defined as
\[
    \sigma_{\text{GK}}^2 = \frac{1}{2} \left( \log\left( \frac{H}{L} \right) \right)^2 - \left( 2\log(2) - 1 \right) \left( \log\left( \frac{C}{O} \right) \right)^2,
\]
where $O$, $H$, $L$, and $C$ denote the opening, high, low, and closing prices, respectively. This formula assumes zero drift and the absence of overnight jumps, making it well suited for capturing intraday risk characteristics.

Figure~\ref{fig:nifty-panel}(e) shows the scatter plot of empirical volatility against Garman-Klass volatility. Figure~\ref{fig:nifty-panel}(f) presents the same comparison on a logarithmic scale. The correlation between the logarithms of empirical and Garman-Klass intra-day volatility is approximately 0.71, suggesting a strong association in the log scale, and highlighting consistency between the two methods of volatility estimation.

\begin{figure}
    \centering
    \includegraphics[width=0.31\linewidth]{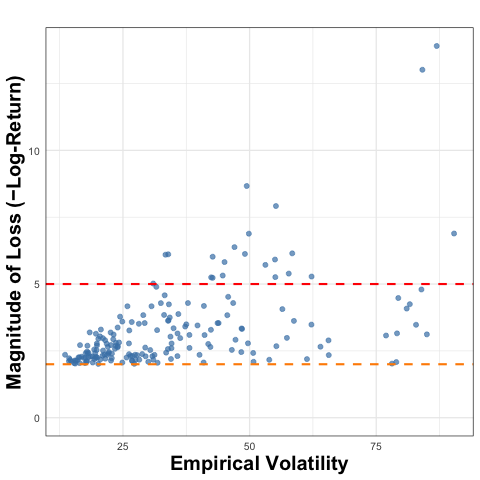}
    \includegraphics[width=0.31\linewidth]{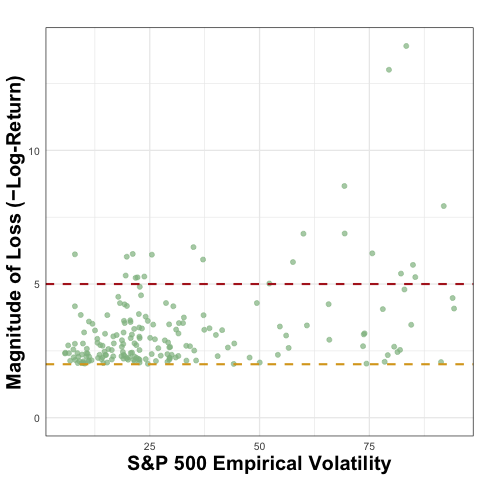}
    \includegraphics[width=0.31\linewidth]{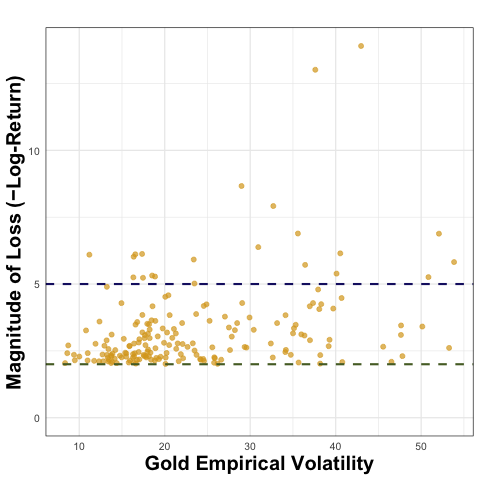}\\[0.5ex]
    \small
    \parbox{0.31\linewidth}{\centering (a)}%
    \parbox{0.31\linewidth}{\centering (b)}%
    \parbox{0.31\linewidth}{\centering (c)}
    
    \caption{Relationship between large losses (negative returns) in the Nifty 50 and empirical volatility measures: (a) own volatility of Nifty 50, (b) S\&P 500 volatility, and (c) gold volatility. Dashed lines indicate thresholds at 2\% and 5\% loss levels.}
    \label{fig:large_drops_vs_vol}
\end{figure}

In addition, we incorporate the S\&P 500 index and gold prices into our analysis to capture global market volatility and assess its influence on the Indian stock market. To study the tail behaviour of returns, we focus on large daily losses in the Nifty 50 index, specifically, those drops, exceeding $2\%$ level, and examine their relationship with daily empirical volatility. Figure~\ref{fig:large_drops_vs_vol} visualises this association across three dimensions of volatility: panel (a) presents the Nifty's own empirical volatility, panel (b) shows S\&P 500 volatility, and panel (c) depicts gold volatility. The plots reveal that large negative returns often coincide with elevated volatility, both domestically and globally. The strong alignment with S\&P 500 volatility suggests the presence of spillover effects from global equity markets, while the relationship with gold volatility may reflect investor flight-to-safety behaviour during periods of financial stress.

\section{Methodology}\label{Sec_Method}

\noindent Suppose $\mathcal{D} = \{(y_i, \mathbf{x}_i) \mid i = 1, 2, \cdots, n\}$ is the observed dataset, where $y_i$ follows a Generalised Pareto Distribution (GPD), i.e.,
\[
y_i \sim \text{GPD}(\mu, \sigma_i, \xi), \quad i = 1, 2, \cdots, n,
\]
with probability density function:
\[
f(y_i \mid \sigma_i, \xi) = \frac{1}{\sigma_i} \left(1 + \xi \frac{y_i - \mu}{\sigma_i} \right)^{-1/\xi - 1},
\]
and $\mathbf{x}_i = \{x_{i1}, x_{i2}, \cdots, x_{ip}\}$ are the covariates associated with the $i^{\text{th}}$ observation. The threshold $\mu$ is assumed known; the shape parameter $\xi$ is same across observations, and the scale parameter $\sigma_i$ is modelled as a log-linear function of covariates:
\[
\log(\sigma_i) = \mathbf{x}_i^\top \boldsymbol{\beta}.
\]
The support of $Y_i$ is $(\mu, \infty)$ if $\xi < 0$, and $(\mu, \mu + \sigma_i / \xi)$ if $\xi > 0$. The special case $\xi = 0$ corresponds to the exponential distribution with mean $\sigma_i$, interpreted as the limit as $\xi \to 0$.
The survival function of the GPD regression model is:
\begin{eqnarray*}
\mathbb{P}(Y > y_0 \mid Y > \mu) &=& \left( 1 - \frac{\xi(y_0 - \mu)}{\exp(\mathbf{x}_i^\top \boldsymbol{\beta})} \right)^{1/\xi} \\
&=& \left(1 + \frac{\xi(y_i - \mu)}{\exp(\mathbf{x}_i^\top \boldsymbol{\beta})} \right)^{-1/\xi}.
\end{eqnarray*}
In financial applications, if $Y$ represents the daily negative return (in percentage), and a market crash is defined as a daily drop exceeding $y_0 = 5\%$, then this survival function quantifies the conditional probability of a crash, given that a drop of, say, $\mu = 2\%$ has already occurred.
The conditional expectation of $Y_i$ given the covariates is:
\[
\mathbb{E}(Y_i\mid Y_i \geq \mu) = 
\begin{cases}
\infty, & \xi \geq 1, \\
\mu + \dfrac{\exp(\mathbf{x}_i^\top \boldsymbol{\beta})}{1 - \xi}, & \xi < 1.
\end{cases}
\]
The variance exists only when $\xi < 0.5$ and is given by:
\[
\mathbb{V}ar(Y_i\mid Y_i \geq \mu) = \dfrac{\exp(2\mathbf{x}_i^\top \boldsymbol{\beta})}{(1 - \xi)^2(1 - 2\xi)}.
\]
This characterisation highlights the influence of covariates on the first two moments of the tail distribution and emphasises the model's ability to capture heavy-tailed behaviour.

\subsection{Log-Likelihood Function}

\noindent The log-likelihood function of the GPD regression model is:
\[
\log L(\xi, \boldsymbol{\beta}) = -\sum_{i=1}^n \mathbf{x}_i^\top \boldsymbol{\beta} + \left(\frac{1}{\xi} - 1\right) \sum_{i=1}^n \log \left[1 + \xi \frac{y_i - \mu}{\exp(\mathbf{x}_i^\top \boldsymbol{\beta})} \right].
\]
This formulation treats the exceedances $y_i > \mu$ as GPD-distributed with covariate-dependent scale. The shape parameter $\xi$ governs the heaviness of the tail, while the covariates modulate the scale of exceedance. In particular, the conditional mean:
\[
\mathbb{E}(Y_i \mid Y_i \geq \mu) = \mu + \frac{\exp(\mathbf{x}_i^\top \boldsymbol{\beta})}{1 - \xi}
\]
exists only for $\xi < 1$, and the variance exists only if $\xi < 0.5$. This structure provides a coherent framework for modelling the magnitude of extreme outcomes while adjusting for covariate information. As shown in \citet{smith1985mle}, the maximum likelihood estimator (MLE) exists asymptotically when $\xi < 1$ and is consistent, asymptotically normal, and efficient provided $\xi < 0.5$.

\subsection{Truncated Cauchy Prior on $\xi$}

\noindent The \texttt{truncated Cauchy(0,1)} prior truncated from above at 1, i.e., the support is $\xi < 1$. The \emph{probability density function} of the standard Cauchy distribution is:
$$
p(\xi) = \frac{1}{\pi(1 + \xi^2)}, \quad \xi \in (-\infty, \infty).
$$
We define the \emph{truncated density}:
$$
p_{\text{trunc}}(\xi) = \frac{1}{Z} \cdot \frac{1}{\pi(1 + \xi^2)}, \quad \text{for } \xi < 1
$$
where $Z$ is the normalising constant, i.e. the total probability over the truncated domain:
\begin{eqnarray}
Z &=& \int_{-\infty}^{1} \frac{1}{\pi(1 + \xi^2)} \, d\xi = \frac{1}{\pi} \cdot \left[ \tan^{-1}(\xi) \right]_{-\infty}^{1}
= \frac{1}{\pi} \cdot \left( \tan^{-1}(1) - \tan^{-1}(-\infty) \right) \nonumber \\
&=&\frac{1}{\pi} \left( \frac{\pi}{4} - (-\frac{\pi}{2}) \right) = \frac{1}{\pi} \cdot \frac{3\pi}{4} = \frac{3}{4}
\end{eqnarray}
Now plug in $Z = \frac{3}{4}$:
$$
p_{\text{trunc}}(x) = \frac{4}{3\pi(1 + \xi^2)}, \quad \xi < 1.
$$
The pdf of \texttt{truncated Cauchy(0,1)} prior density, with upper truncation at $\xi = 1$.
$$
p(\xi) = \frac{4}{3\pi(1 + \xi^2)}, \quad \text{for } \xi < 1
$$
% $$
% \boxed{
% p(\xi) = \frac{4}{3\pi(1 + \xi^2)}, \quad \text{for } \xi < 1
% }
% $$

\subsection{\texorpdfstring{Cauchy Prior on $\boldsymbol{\beta}$}{Cauchy Prior on beta}}

In our Bayesian regression framework, we assign a standard Cauchy prior to the regression coefficients $\boldsymbol{\beta}$, see \cite{gelman2008cauchy}. The Cauchy distribution is symmetric and heavy-tailed, making it well-suited for scenarios where large effect sizes are plausible or when robust shrinkage is needed. The standard Cauchy prior has the following probability density function:
\[
p(\beta) = \frac{1}{\pi (1 + \beta^2)}, \quad \beta \in \mathbb{R}.
\]
A key property of the Cauchy distribution is that it lacks a finite mean, variance, and higher moments. This feature makes it a particularly attractive choice as a weakly informative prior, see \cite{berger1985statistical}. It imposes minimal structure on the parameter space, allowing the data to dominate the posterior inference, while still penalising extremely large values less severely than priors with finite variance, such as the Gaussian. Consequently, the Cauchy prior provides a balance between regularisation and flexibility, making it suitable for sparse or high-dimensional regression models where robustness and parsimony are desired.

\subsection{\texorpdfstring{Lasso Prior on $\boldsymbol{\beta}$}{Lasso Prior on beta}}

An alternative to the Cauchy prior is the Lasso prior, which corresponds to a Laplace (double-exponential) distribution on the regression coefficients $\boldsymbol{\beta}$, see \cite{hastie2009elements}, \cite{park2008bayesianlasso}. The Lasso prior encourages sparsity in the estimated coefficients by applying stronger shrinkage towards zero, making it particularly useful for high-dimensional models or when variable selection is of interest. The probability density function of the Laplace distribution with scale parameter $\lambda > 0$ is given by:
\[
p(\beta) = \frac{\lambda}{2} \exp(-\lambda |\beta|), \quad \beta \in \mathbb{R}.
\]
This prior has a sharp peak at zero and heavier tails than the Gaussian, enabling it to shrink small coefficients more aggressively while allowing larger coefficients to remain relatively unaffected. The result is a sparse posterior mode, with many coefficients estimated as exactly zero, aligning with the behaviour of the classical Lasso estimator. Unlike the Cauchy prior, the Laplace distribution has finite mean and variance, making it more suitable when a moderate degree of regularisation is desired without the extreme heavy tails of the Cauchy prior. Its effectiveness in automatic variable selection and computational tractability makes the Lasso prior a popular choice in Bayesian sparse regression models.

\subsection{\texorpdfstring{Ridge Prior on $\boldsymbol{\beta}$}{Ridge Prior on beta}}

The Ridge prior assumes a Gaussian (normal) distribution on the regression coefficients $\boldsymbol{\beta}$, offering smooth shrinkage without inducing sparsity, \cite{hastie2009elements}. It is commonly used when multicollinearity is present or when all predictors are believed to have small, non-zero effects. The probability density function of the Ridge prior with precision parameter $\tau$ is:
\[
p(\beta) = \sqrt{\frac{\tau}{2\pi}} \exp\left(-\frac{\tau}{2} \beta^2\right), \quad \beta \in \mathbb{R}.
\]
This is equivalent to placing an independent $\mathcal{N}(0, \tau^{-1})$ prior on each coefficient $\beta$. The Ridge prior results in posterior estimates that are biased toward zero, but unlike the Lasso prior, it does not set any coefficients exactly to zero. Consequently, it is well-suited for settings where all predictors contribute weakly to the response.

In the Bayesian framework, the Ridge prior leads to a conjugate posterior when combined with a Gaussian likelihood, allowing for closed-form posterior inference in linear models. Its computational simplicity and regularising effect make it a widely used choice in models where interpretability through sparsity is not essential but stabilisation of estimates is desired.

\subsection{\texorpdfstring{Zellner's g-Prior on $\boldsymbol{\beta}$}{Zellner's g-Prior on beta}}

Zellner's \textit{g}-prior is a popular conjugate prior used in Bayesian linear regression, particularly when the design matrix \(\mathbf{X}\) is fixed and known, see \cite{zellner1986gprior},\cite{bove2011hyper}. It imposes a multivariate normal prior on the regression coefficients \(\boldsymbol{\beta}\), with the covariance structure informed by the design matrix:
\[
\boldsymbol{\beta}  \sim \mathcal{N}\left( \mathbf{0},\ g  (\mathbf{X}^\top \mathbf{X})^{-1} \right),
\]
where \(g > 0\) is a scalar hyperparameter controlling the strength of the prior relative to the likelihood. This prior has several attractive properties:
\begin{itemize}
  \item It is invariant under linear transformations of the design matrix.
  \item The posterior mean shrinks toward zero as \(g \to 0\), while as \(g \to \infty\), the prior becomes non-informative.
  \item The use of \((\mathbf{X}^\top \mathbf{X})^{-1}\) ensures the prior reflects the geometry of the predictors.
\end{itemize}

Zellner's \textit{g}-prior is especially useful for Bayesian model comparison and variable selection, as it facilitates closed-form expressions for marginal likelihoods and Bayes factors. However, care must be taken in the choice of \(g\), as it significantly influences the inference. Empirical Bayes, fixed \(g\), or hierarchical models placing a hyperprior on \(g\) are common approaches to address this.

\subsection{MAP Estimation for Efficient Bayesian Model Exploration}

Maximum a Posteriori (MAP) estimation provides a computationally efficient alternative to full Bayesian inference, especially when exploring different model specifications or prior structures. Unlike posterior summaries such as the posterior mean or credible intervals that rely on Markov Chain Monte Carlo (MCMC) methods, MAP estimation avoids the computational burden of sampling by framing inference as an optimisation problem. Formally, the MAP estimate is defined as:
\[
\hat{\theta}_{\text{MAP}} = \arg\max_{\theta} \, p(\theta \mid \mathcal{D}) = \arg\max_{\theta} \left[ \log p(\mathcal{D} \mid \theta) + \log p(\theta) \right],
\]
where \( \mathcal{D} \) denotes the observed data, \( p(\mathcal{D} \mid \theta) \) is the likelihood function, and \( p(\theta) \) is the prior distribution. In our case $\theta = (\boldsymbol{\beta},\xi)$. This above expression highlights that MAP estimation balances the fit to the data and the influence of prior information, analogous to penalised likelihood methods in the frequentist setting.

MAP estimation is often preferred in high-dimensional problems or when computational resources are limited, as it offers faster and more stable inference than MCMC, which can suffer from slow convergence and poor mixing. Moreover, it enables quick evaluation of multiple models or regularisation schemes, such as Lasso or Ridge, where priors play the role of sparsity-inducing penalties.

Importantly, MAP estimation is well-suited for the model exploration phase. Once a final model is selected, based on predictive performance or interpretability, one may then employ MCMC on that specific model to obtain a full characterisation of the posterior distribution. In this way, MAP provides an efficient screening tool, while full Bayesian inference is reserved for the final model of interest.

\section{Simulation Stusy of Frequentist Properties}\label{Sec_Simulation}

To evaluate the performance of different regularised Generalised Pareto Distribution (GPD) regression models, we conduct a simulation study comparing four Bayesian prior structures on the regression coefficients: the \textbf{Cauchy prior}, \textbf{Lasso} (\(\ell_1\) penalty), \textbf{Ridge} (\(\ell_2\) penalty), and \textbf{Zellner's g-prior}. Our aim is to assess and compare their predictive accuracy, estimation error, model complexity (via AIC and BIC), and computational efficiency.

We simulate \( N = 100 \) datasets, each consisting of \( n = 100 \) observations. For each dataset, the covariate matrix \( X \in \mathbb{R}^{n \times p} \) is generated from a multivariate normal distribution with identity covariance, where the number of predictors is \( p = 5 \). The true regression coefficients \( \beta \) are sampled from a standard normal distribution, and the GPD shape parameter \( \xi \) is drawn from a uniform distribution on the interval \((-0.5, 0.5)\). The response variable \( y \) is generated from a GPD model with a fixed location parameter \( \mu = 2 \) and a scale parameter \( \sigma = \exp(X\beta) \). Each dataset is then split into a training set (80\%) and a testing set (20\%).

Each of the four models is estimated using maximum a posteriori (MAP) estimation via the \texttt{BFGS} optimisation algorithm. The Cauchy prior is specified as a weakly informative heavy-tailed prior on the coefficients \( \beta \), along with a truncated Cauchy prior on the shape parameter \( \xi \). The Lasso model imposes an \(\ell_1\) penalty on the regression coefficients, while the Ridge model uses an \(\ell_2\) penalty; in both cases, the regularisation parameter is chosen using 5-fold cross-validation. The Zellner’s g-prior assumes a prior of the form \( \beta \sim \mathcal{N}(0, g (X^\top X)^{-1}) \), and the hyperparameter \( g \) is selected through cross-validation.

We assess model performance using several metrics. These include the root mean squared error (RMSE) of predictions on the test data, as well as the RMSE for recovering the true coefficient vector \( \beta \) and the shape parameter \( \xi \). We also compute the Akaike Information Criterion (AIC) and the Bayesian Information Criterion (BIC), both based on the negative log-likelihood (excluding the contribution of the prior). In addition, we record the computational time taken to fit each model, and express it as a multiple relative to the time taken for the Cauchy prior model, which serves as the baseline.

\begin{table}[ht]
\centering
\caption{Simulation comparison of GPD regression models with different priors}
\label{tab:gpd-simulation-summary}
\vspace{0.5em}
\renewcommand{\arraystretch}{1.3} % Increases row height
\begin{tabular}{l|c|c|c|c}
\hline
\textbf{Metric} & \textbf{Cauchy} & \textbf{Lasso} & \textbf{Ridge} & \textbf{g-prior} \\
\hline
RMSE (y)        & 11.36 & 10.37 & 10.65 & 10.43 \\
RMSE ($\beta$)  & 0.10  & 0.16  & 0.16  & 0.18  \\
RMSE ($\xi$)    & 0.08  & 0.08  & 0.09  & 0.11  \\
AIC             & 162.67 & 177.15 & 175.74 & 180.94 \\
BIC             & 176.97 & 191.44 & 190.03 & 195.24 \\
Time (sec)      & 0.00  & 0.60  & 0.53  & 8.45  \\
Time (relative) & 1.00  & 175.64 & 155.69 & 2476.86 \\
\hline
\end{tabular}
\end{table}

Finally, the results across the 100 simulations are aggregated using the median to ensure robustness. The outcome is summarised in the Table~\ref{tab:gpd-simulation-summary}, displaying predictive accuracy, parameter estimation error, model complexity, and computational cost for each prior, facilitating a comprehensive evaluation of their relative performance.

\section{Empirical Study of Market Crashes}\label{Sec_Empirical}

In this section, we investigate the application of Generalised Pareto Distribution (GPD) regression for modelling extreme negative returns in the Indian equity market. Our analysis is based on daily returns of the NSE Nifty 50 index, augmented with realised volatility measures derived from domestic and global financial assets. The central objective is to assess the predictive performance and interpretability of GPD regression models under a range of prior assumptions on the regression coefficients, within a Bayesian framework for financial tail risk modelling.

\subsection*{Data Description and Preprocessing}

We work with a cleaned dataset comprising daily log-returns of the NSE Nifty 50 index, with all missing entries removed. Denoting the daily return at time $t$ as $r_t$, we extract a subset of observations where $r_t < -2\%$, corresponding to the left tail of the return distribution. This subset constitutes approximately 4.6\% of all trading days. Among these tail events, 11.6\% show losses exceeding 5\%, indicating the presence of disproportionately large market crashes within the extreme left tail.

To facilitate interpretation and estimation, negative returns are transformed into absolute values, i.e., $y_t = |r_t|$ for $r_t < -2\%$. We construct covariates based on (i) empirical volatility and (ii) Garman--Klass (GK) volatility for the Nifty 50, S\&P 500, and gold. These volatility variables are standardised to have mean zero and unit variance. The baseline model includes only the intercept and Nifty empirical volatility, while richer covariate sets are examined in subsequent analyses.

% \subsection*{Model Specification}

% The tail losses $y_i$ are assumed to follow a GPD conditional on covariates $x_i$, given by:
% \[
% y_i \mid x_i, y_i > \mu \sim \text{GPD}(\mu, \sigma(x_i), \xi), \quad \text{where} \quad \log \sigma(x_i) = x_i^\top \beta,
% \]
% with a fixed threshold $\mu = 2$, regression coefficients $\beta \in \mathbb{R}^p$, and shape parameter $\xi$. The log-link function ensures positivity of the scale parameter $\sigma(x_i)$. We estimate model parameters via maximum a posteriori (MAP) optimisation under four prior assumptions on $\beta$:

% \begin{enumerate}
%     \item \textbf{Cauchy prior:} a heavy-tailed weakly informative prior;
%     \item \textbf{Lasso prior:} a Laplace prior inducing sparsity via $L_1$ penalisation;
%     \item \textbf{Ridge prior:} a Gaussian prior corresponding to $L_2$ regularisation;
%     \item \textbf{Zellner's $g$-prior:} an informative prior with data-adaptive covariance structure.
% \end{enumerate}

% The regularisation hyperparameters (penalty weight or $g$) are chosen via 5-fold cross-validation using training data.

\subsection*{Predictive Evaluation and Model Comparison}

We randomly split the dataset into training (80\%) and test (20\%) sets. Out-of-sample predictive performance is assessed via Root Mean Squared Error (RMSE), and in-sample fit is evaluated using Akaike Information Criterion (AIC) and Bayesian Information Criterion (BIC), both computed from the negative log-likelihood (excluding prior terms). For the Lasso model, degrees of freedom are adjusted to reflect the number of non-zero coefficients.

\begin{table}[ht]
\centering
\caption{Model evaluation using RMSE, AIC, and BIC.}
\label{tab:emp_model-eval}
\vspace{0.5em}
\renewcommand{\arraystretch}{1.3}
\begin{tabular}{lccc}
\hline
\textbf{Prior Models} & \textbf{RMSE} & \textbf{AIC} & \textbf{BIC} \\
\hline
Cauchy     & 1.58 & 298.06 & 322.20 \\
Lasso      & 1.80 & 311.57 & 335.71 \\
Ridge      & 1.73 & 302.07 & 326.20 \\
Zellner's $g$-prior & 1.94 & 308.12 & 332.25 \\
\hline
\end{tabular}
\end{table}

Among all models, the Cauchy prior delivers the best fit, achieving the lowest AIC and BIC values, along with the smallest RMSE, suggesting strong predictive accuracy and robust in-sample performance. The Ridge and $g$-prior models provide competitive alternatives, whereas the Lasso model performs relatively poorly, possibly due to over-shrinkage.

To understand the economic implications of volatility on tail risk, we compute both the conditional expectation $\mathbb{E}[Y \mid Y > \mu, x]$ and the conditional exceedance probability $\mathbb{P}(Y > 5 \mid Y > 2, x)$ under the fitted Cauchy model. Results show a nonlinear increase in both quantities with rising volatility. At the lower end (e.g., 10th percentile of volatility), expected losses are around 2.4 with tail event probabilities below 5\%. At the upper end (90th percentile), expected losses exceed 7 with crash probabilities above 65\%.

\begin{figure}[ht]
    \centering
    \begin{minipage}[b]{0.32\linewidth}
        \centering
        \includegraphics[width=\linewidth]{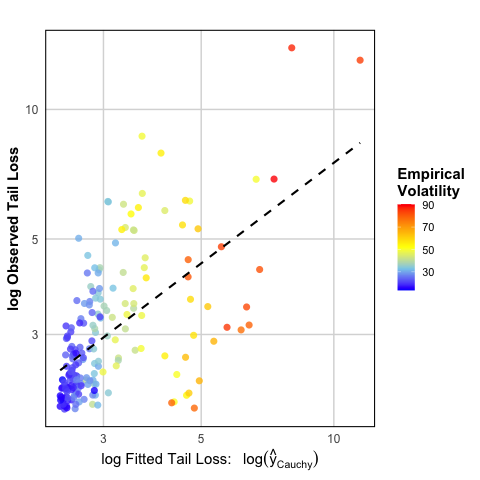}
        \vspace{0.3em}
        
        (a) %Fitted vs Observed Tail Loss
    \end{minipage}
    \hfill
    \begin{minipage}[b]{0.32\linewidth}
        \centering
        \includegraphics[width=\linewidth]{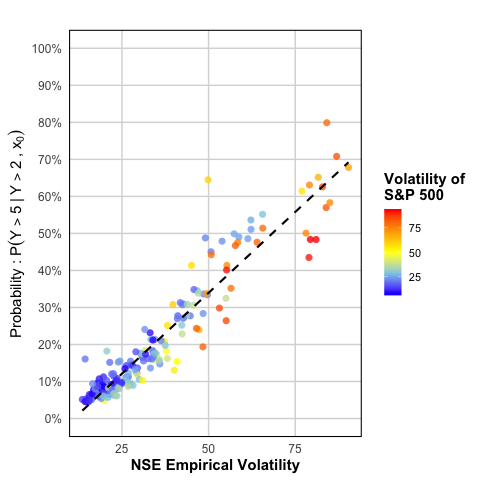}
        \vspace{0.3em}
        
        (b) %Tail Probability vs Nifty Volatility (coloured by S\&P 500)
    \end{minipage}
    \hfill
    \begin{minipage}[b]{0.32\linewidth}
        \centering
        \includegraphics[width=\linewidth]{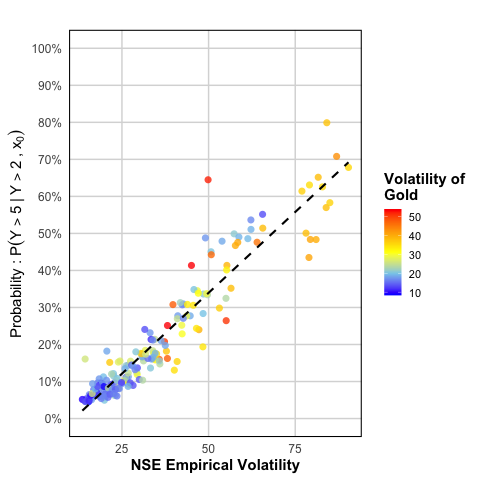}
        \vspace{0.3em}
        
        (c) %Tail Probability vs Nifty Volatility (coloured by Gold)
    \end{minipage}
    
    \caption{Relationship between extreme tail events in the Nifty 50 index and market volatility. Panel (a) compares the log-fitted tail losses from the GPD model under a Cauchy prior with actual observed losses. Panels (b) and (c) show the model-based conditional tail probabilities as functions of Nifty volatility, coloured by S\&P 500 and gold volatilities, respectively.}
    \label{fig:tail-risk-volatility}
\end{figure}

Figure~\ref{fig:tail-risk-volatility} presents three visual diagnostics. Panel (a) plots the fitted vs. observed tail losses on the log scale, with colour indicating empirical Nifty volatility. The upward trend and colour gradient affirm that higher volatility coincides with more severe losses, though dispersion increases in the tail. Panels (b) and (c) display the fitted probability $\mathbb{P}(Y > 5 \mid Y > 2, x_0)$ as a function of Nifty volatility, coloured by the S\&P 500 and gold volatilities, respectively. \emph{These plots reveal clear global spillover effects: both S\&P 500 and gold volatilities enhance the likelihood of extreme losses in the Indian market, the former via market correlation and the latter via flight-to-safety dynamics.}

%\subsection*{Summary}

This empirical exercise highlights that the tail behaviour of Indian equity returns is jointly shaped by local volatility conditions and global financial signals. The GPD model with a Cauchy prior not only provides superior predictive performance but also captures the nonlinear and interactive effects of volatility on extreme losses. These findings underscore the importance of robust regularisation and the inclusion of international covariates and global spill-over effect in tail risk modelling of Indian stock market.

\section{Conclusion}\label{Sec:Conclusion}

In this study, we developed and applied a Bayesian Generalised Pareto Regression (GPR) framework to model and predict extreme negative returns in the Indian equity market, particularly focusing on the Nifty 50 index. The core innovation lies in modelling the scale parameter of the Generalised Pareto Distribution as a log-linear function of market volatility indicators, while keeping the shape parameter same for all data points. This allows the model to adapt to changing volatility regimes while preserving interpretability and parsimony.

Our theoretical development was supported by a thorough simulation study comparing different regularisation strategies, including Cauchy, Lasso, Ridge, and Zellner’s $g$-prior; within a MAP estimation framework. Among these, the Cauchy prior emerged as the most effective in balancing predictive accuracy and model complexity. This observation was further validated in the empirical analysis of actual market crash events, where the Cauchy Prior achieved the lowest AIC, BIC, and RMSE values.

Empirical findings highlight a strong nonlinear association between volatility and the likelihood and magnitude of extreme losses. We demonstrated that both domestic (Nifty) and global (S\&P 500, gold) volatility measures significantly influence the tail risk. The inclusion of global covariates like S\&P 500 and gold volatilities proved crucial in capturing spillover and flight-to-safety effects during periods of financial stress.

Visual diagnostics revealed that in high-volatility regimes, the conditional probability of a crash; defined as a loss exceeding 5\% given a 2\% drop; can exceed 60\%, with expected losses rising sharply. These insights underscore the importance of incorporating volatility-sensitive covariates and flexible modelling strategies when forecasting rare but impactful financial events.

Overall, our study demonstrates that Bayesian GPD regression, particularly under heavy-tailed priors like the Cauchy, provides a powerful and interpretable tool for tail risk modelling in financial markets. It holds promise for future applications in systemic risk monitoring, stress testing, and portfolio tail risk management. Future research could explore dynamic extensions, integrate macroeconomic signals, and investigate hierarchical Bayesian formulations that allow for time-varying shape parameters or latent volatility drivers.

% \section*{Acknowledgements}
% I am indeed grateful to the Editors for their guidance and counsel. I am very grateful to the reviewer for valuable comments and suggestions of generously listing many useful references. 
% \vskip0.3cm

% The author do not have any financial or non-financial conflict of interest to declare for the research work included in this article.

%%%%%%%%%%%%%%%%%    REFERENCES    %%%%%%%%%%%%%%%%%%%%%%%%%%
\interlinepenalty  = 10000  % optional, used to move reference lines to next page
\bibliographystyle{plainnat}
\bibliography{biblio_ref}

\end{document}